\documentclass{Rinton-P9x6}

\usepackage{bm}
\usepackage{latexsym}
\usepackage{amsmath}
\usepackage{amsfonts}
\usepackage{amssymb}

\newcommand{\ket}[1]{\mbox{$|#1\rangle$}}
\newcommand{\bra}[1]{\mbox{$\langle#1|$}}
\def\identity{\leavevmode\hbox{\small1\kern-3.2pt\normalsize1}}%

\begin{document}

\title{Decoherence in a quantum walk on the line}

\author{Viv Kendon}
\author{Ben Tregenna}

\address{Optics Section, Blackett Laboratory, Imperial College, London, SW7 2BW, UK
\\E-mail: Viv.Kendon@ic.ac.uk
}


\maketitle

\abstracts{
We have studied how decoherence affects a quantum walk on the line.
As expected, it is highly sensitive, consisting as it does of an
extremely delocalized particle.  We obtain an expression for
the rate at which the standard deviation falls from the quantum
value as decoherence increases and show that it is proportional
to the number of decoherence ``events'' occuring during the walk.
}

\section{Introduction}

Quantum walks on discrete lattices are of interest to the quantum
information community because classical random walks underpin
important classical algorithms.  The aim is to find more
powerful quantum counterparts.  First it is useful to understand
the basic properties of quantum walks on simple lattices.
Two versions have been studied, continuous time\cite{farhi98a}
and discrete-time coined\cite{aharonov92a} quantum walks.
Here we consider only the coined version.
Coined quantum walks on a line, cycle, and the hypercube have been solved
exactly\cite{ambainis01a,AAKV00a,moore01a}, and numerical studies used to
explore quantum walks on two dimensional lattices\cite{mackay01a,tregenna02a}.
The quantum walks show a speed up over the equivalent classical random
walk, a quadratic increase in standard deviation for a
walk on a line, a quadratically faster mixing time of a walk on a cycle and
an exponential increase in the hitting time to the opposite corner
for a walk on a hypercube\cite{kempe02a}.

\section{Qauntum walk on a line}

A quantum coin is a
$d$-dimensional quantum system which is ``flipped'' by applying
a unitary operator $\mathbf{C}$.  For a quantum walk on a line,
$d=2$ (``left'' and ``right'') and without loss of generality
$\mathbf{C}$ can be taken to be a Hadamard ($\pi/2$) rotation.
The particle then moves in a superposition of the two possible
directions indicated by the coin, $\mathbf{S}\ket{x,a} = \ket{x+a,a}$,
where $x$ is the particle position and $a\in\{-1,+1\}$ is the state of the coin.
The unitary operator for each step of the walk is thus
$\mathbf{U}=\mathbf{S}\cdot(\mathbf{C}\otimes\mathbb{I})$.
The moments have been calculated\cite{ambainis01a},
for asymptotically large times $T$ for a walk starting at the origin,
\begin{equation}
\langle x^2\rangle = (1-1/\sqrt{2})T^2 = \sigma^2(T),
\label{indepmoments}
\end{equation}
independent of initial coin state, and
\begin{equation}\label{depmoments}
\langle x \rangle_a = a(1-1/\sqrt{2})T,
\end{equation}
where $a\in\{1,-1\}$ is the initial coin state.
The standard deviation (from the origin) $\sigma(T)$
is thus linear in $T$, in contrast to $\sqrt{T}$ for the classical walk.

\section{Decoherence in the walk on a line}

To model decoherence in this system we use a discrete master equation 
\begin{equation}\label{decme}
\rho(t+1) = (1-p)U\rho(t) U^\dagger + p \sum_i I\!\!P_i U\rho(t)
U^\dagger I\!\!P_i,
\end{equation}
where the summation runs over the dimensions of the Hilbert space on which
the decoherence occurs, either the coin, the particle, or both.
The projectors $I\!\!P_i$ act in the computational basis,
and $p$ is the probability of a decoherence event happening per time step.
This equation was evolved numerically\cite{kendon02a} for various choices of
$I\!\!P$.  All choices produced the same general form for the decay of 
$\sigma(T,p)$ from the quantum ($p=0$) to the classical ($p=1$) value,
with small differences in the rates.
The slope of $\sigma(T,p)$ is finite as $p\rightarrow 0$ and zero
at $p = 1$. Here we calculate $\sigma(T,p)$ analytically
for $pT \ll 1$ and $T \gg 1$ for the case where $I\!\!P$ is the
projector onto the preferred basis $\{\ket{a,x}\}$, i.e., decoherence
affecting both particle and coin.

The probability distribution for finding the
particle in the state \ket{a,x} in the presence of decoherence can be written,
\begin{equation}
P(x,a,T,p) = (1-p)^TP(x,a,T) + p(1-p)^{T-1}P^{(1)}(x,a,T) + \dots ,
\label{PxapTgen}
\end{equation}
where $P(x,a,T)$ is the distribution obtained for a perfect walk and
$P^{(i)}(x,a,T)$ is the sum of all the ways to have exactly $i$ noise
events, e.g.,
\begin{equation}
P^{(1)}(x,a,T) = \sum_{t=1}^T \sum_y\sum_b P(y,b,t) P_{yb}(x,a,T-t),
\label{P1gen}
\end{equation}
where $P_{yb}(x,a,T-t)$ is the distribution obtained by a perfect
(no noise) walk starting in state $\ket{y,b}$ for $T-t$ steps.
For the ideal walk, $\sigma^2(T) \equiv \sum_x\sum_a x^2 P(x,a,T)$,
and for the walk with decoherence,
\begin{equation}
\sigma^2(T,p) \equiv \sum_x\sum_a x^2 P(x,a,T,p).
\label{sddefgen}
\end{equation}
Taking Eqn.~(\ref{PxapTgen}) to first order in $p$,
and substituting along with Eq.~(\ref{P1gen}) into  Eq.~(\ref{sddefgen}) gives
\begin{equation}
\sigma^2(T,p) \simeq \sum_{x,a} x^2 \left\{\mbox{\rule[-1em]{0em}{2.7em}}(1-pT)P(x,a,T) + p\sum_{t=1}^T \sum_{y,b} P(y,b,t) P_{yb}(x,a,T-t)\right\}.
\label{sigTp}
\end{equation}
The first term on the r.h.s. is (by definition) $(1-pT)\sigma^2(T)$.
Noting that $P_{yb}(x,a,T-t)$ is a translation of a walk starting at
the origin, $P_{yb}(x,a,T-t) = P_{0b}(x-y,a,T-t)$.
Relabelling the summed variable $x$ to $(x+y)$ then enables
the sums over $x$ and $a$ to be performed in the second term,
\begin{eqnarray}\label{sdpart}
&& p\sum_{t=1}^T \sum_y\sum_b P(y,b,t)\sum_x\sum_a (x+y)^2 P_{0b}(x,a,T-t) \nonumber\\
& = & p\sum_{t=1}^T \sum_y\sum_b P(y,b,t) \left\{\sigma^2_{0b}(T-t) + 2y\langle x\rangle_{0b}^{(T-t)} + y^2 \right\}.
\end{eqnarray}
From Eqn.~(\ref{indepmoments}), $\sigma^2_{0b}(T-t)$ does not
depend on $b$, so the summation over $y$ and $b$ may be performed trivially.
The summation applied to $y^2$ gives $\sigma^2_0(t)$ by definition.
This leaves only the evaluation of
\begin{equation}
 2p\sum_{t=1}^T \sum_{y,b}P(y,b,t)y\langle x\rangle_{0b}^{(T-t)}
 =  2p(1-1/\sqrt{2})\sum_{t=1}^T (T-t)\sum_{y,b}yb~ P(y,b,t),
\end{equation}
where we have used Eqn.~(\ref{depmoments}).
We note that this term does not depend on whether
the initial coin state is plus or minus one and so we may
include both these possibilities equally.
Also, by the symmetry of the walk, it is possible to rewrite a 
probability function for travelling from state \ket{0,c} to \ket{y,b} in
the reverse order, i.e. as a probability for moving from \ket{y,b} to
\ket{0,c}.  Care must be taken to ensure that the signs of each term
due to the coefficient $yb$ in the summation are maintained.
We obtain
\begin{eqnarray}
&&2p\!\!\sum_{t,y,b}\!\!P(y,b,t)y\langle x\rangle_{0b}^{(T-t)}
= p(1-1/\sqrt{2})\sum_{t=1}^T (T-t)\!\!\sum_{y,b,c}(1-2\delta_{b,c})yb~P_{yb}(0,c,t)\nonumber\\
&&= p(1-1/\sqrt{2})\sum_{t=1}^T
(T-t)\!\!\left[\sum_{y,b,c}yb~P_{y,b}(0,c,t)-2\sum_{y,b}yb~P_{yb}(0,b,t)]\right]\!\!,
\end{eqnarray}
treating the two parts with and without a delta function independently.
Expanding the summations over $b$ and translating
the particle basis by $-y$ gives
\begin{eqnarray}
&&2p\sum_{t=1}^T \sum_{y,b}P(y,b,t)y\langle x\rangle_{0b}^{(T-t)}\nonumber\\
& = &p(1-1/\sqrt{2})\!\!\sum_{t=1}^T(T-t)\!\!\left[\sum_{y,c}y P_{0,-1}(y,c,t)
-4\sum_y y~P_{y,+1}(y,+1,t)\right]\!\!\!.
\label{eq:term2}
\end{eqnarray}
The final summation over $y$ may be bounded above by noting that
\begin{eqnarray}
P_{0-1}(y,-1,t)&=&|\bra{y,-1}U^t\ket{0,-1}|^2\nonumber\\
	       &=&\frac{1}{2}|\bra{y+1,1}U^{t-1}\ket{0,-1} -
	       \bra{y+1,-1}U^{t-1}\ket{0,-1}|^2 \nonumber\\
	       &\leq&\frac{1}{2}\sum_c P_{0,-1}(y+1,c,t-1).
\end{eqnarray}
Using this in Eqn.~(\ref{eq:term2}) gives,
\begin{eqnarray}
2p\sum_{t,y,b}P(y,b,t)y\langle x\rangle_{0b}^{(T-t)}
& \leq & p(1-1/\sqrt{2})\sum_{t=1}^T(T-t) \left[\langle y
    \rangle_{0,-1}^t -2 \langle y \rangle_{0,-1}^{(t-1)}
    +2\right]\nonumber\\
&\leq&p(1-1/\sqrt{2})\sum_{t=1}^T(T-t)\left[(1-1/\sqrt{2})t+\sqrt{2}\right],
\end{eqnarray}
where Eqn.~(\ref{depmoments}) has been used for the average values.
Combining these results in the full expression
for $\sigma^2(T,p)$, Eqns.~(\ref{sigTp}, \ref{sdpart}), and
performing the summations over $t$ using
$\sum t= T(T+1)/2$ and $\sum t^2 = T^3/3 +T^2/2 +T/6$, gives
\begin{equation}
\sigma^2(T,p) \leq \sigma_0^2(T)\left[ 1 - \frac{\sqrt{2}}{6}pT +
p(\sqrt{2}-1)+ \ldots \right].
\end{equation}
Taking the square root gives as an upper bound on the standard deviation,
\begin{equation}
\sigma(T,p) \leq \sigma(T)\left[1-\frac{pT}{6\sqrt{2}} +
\frac{p}{\sqrt{2}}(1-1/\sqrt{2}) + O(p^2, 1/T)\right].
\label{eq:sigpTfin}
\end{equation}

\section{Summary}

\begin{figure}
  \begin{center}
    \resizebox{0.6\columnwidth}{!}{\includegraphics{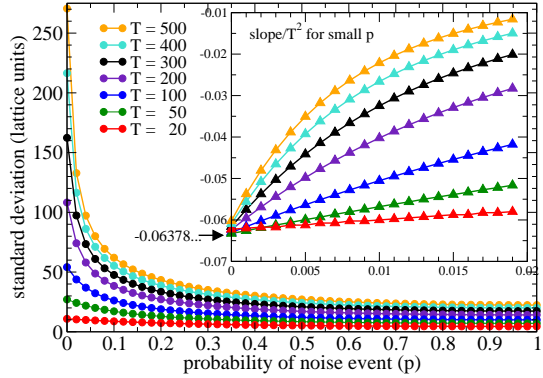}}
    \caption{Standard deviation $\sigma(T,p)$ of the particle position
		on the line for $T=20$--500.
		Inset shows small $p$ region scaled by $T^{-2}$ converging on
		value of $0.06378\dots$.}
    \label{fig:dec-both}
  \end{center}
\end{figure}
Equation (\ref{eq:sigpTfin})  compares well with simulation data, once a second order correction
for $\sigma(T) = (1-1/\sqrt{2})^{1/2}(T-1/T)$ is taken into
account. The $1/T$ form of this correction was found in 
Ambainis et.~al\cite{ambainis01a},
however, we determined the coefficient $(1-1/\sqrt{2})^{1/2}$ numerically
for a wide range of evolution times $T$. The bounding procedure
applied here is reasonably accurate, numerical studies give the
coefficent of $p$ in the above expansion as 0.09566, compared with the
bound of 0.20711.
The first order dependence is thus proportional to $pT$, 
the number of decoherence events during the whole quantum walk.
For a given decoherence rate $p$, 
the standard deviation initially decreases linearly in $T$.
Related results for decoherence on the coin only can be found
in Brun et.~al\cite{brun02a}.

\section*{Acknowledgments}

We thank Peter Knight, Will Flanagan, Rik Maile, Julia Kempe, Jens Eisert, Todd Brun, and Hilary Carteret for useful discussions.
Funded by the UK Engineering and Physical Sciences Research Council.


\begin{thebibliography}{99}
\expandafter\ifx\csname bibnamefont\endcsname\relax
  \def\bibnamefont#1{#1}\fi
\expandafter\ifx\csname bibfnamefont\endcsname\relax
  \def\bibfnamefont#1{#1}\fi
\expandafter\ifx\csname url\endcsname\relax
  \def\url#1{\texttt{#1}}\fi
\expandafter\ifx\csname urlprefix\endcsname\relax\def\urlprefix{URL }\fi
\providecommand{\bibinfo}[2]{#2}
\providecommand{\eprint}[2][]{\url{#2}}

\bibitem{farhi98a}
\bibinfo{author}{\bibfnamefont{E.}~\bibnamefont{Farhi}} \bibnamefont{and}
  \bibinfo{author}{\bibfnamefont{S.}~\bibnamefont{Gutmann}},
  \bibinfo{journal}{Phys.~Rev.~A} \textbf{\bibinfo{volume}{58}},
  \bibinfo{pages}{915--928} (\bibinfo{year}{1998}), \eprint{quant-ph/9706062}.

\bibitem{aharonov92a}
\bibinfo{author}{\bibfnamefont{Y.}~\bibnamefont{Aharonov}},
  \bibinfo{author}{\bibfnamefont{L.}~\bibnamefont{Davidovich}},
  \bibnamefont{and} \bibinfo{author}{\bibfnamefont{N.}~\bibnamefont{Zagury}},
  \bibinfo{journal}{Phys. Rev. A}
  \textbf{\bibinfo{volume}{48}}(\bibinfo{number}{2}), \bibinfo{pages}{1687}
  (\bibinfo{year}{1992}).

\bibitem{ambainis01a}
\bibinfo{author}{\bibfnamefont{A.}~\bibnamefont{Ambainis}},
  \bibinfo{author}{\bibfnamefont{E.}~\bibnamefont{Bach}},
  \bibinfo{author}{\bibfnamefont{A.}~\bibnamefont{Nayak}},
  \bibinfo{author}{\bibfnamefont{A.}~\bibnamefont{Vishwanath}},
  \bibnamefont{and} \bibinfo{author}{\bibfnamefont{J.}~\bibnamefont{Watrous}},
  \bibinfo{journal}{Proc. 33rd STOC} pp. \bibinfo{pages}{60--69}
  (\bibinfo{year}{2001}).

\bibitem{AAKV00a}
\bibinfo{author}{\bibfnamefont{D.}~\bibnamefont{Aharanov}},
  \bibinfo{author}{\bibfnamefont{A.}~\bibnamefont{Ambainis}},
  \bibinfo{author}{\bibfnamefont{J.}~\bibnamefont{Kempe}}, \bibnamefont{and}
  \bibinfo{author}{\bibfnamefont{U.}~\bibnamefont{Vazirani}},
  \bibinfo{journal}{Proc. 33rd STOC} pp. \bibinfo{pages}{50--59}
  (\bibinfo{year}{2001}), \eprint{quant-ph/0012090}.

\bibitem{moore01a}
\bibinfo{author}{\bibfnamefont{C.}~\bibnamefont{Moore}} \bibnamefont{and}
  \bibinfo{author}{\bibfnamefont{A.}~\bibnamefont{Russell}},
  \emph{\bibinfo{title}{Quantum walks on the hypercube}}
  (\bibinfo{year}{2001}), \eprint{quant-ph/0104137}.

\bibitem{mackay01a}
\bibinfo{author}{\bibfnamefont{T.~D.} \bibnamefont{Mackay}},
  \bibinfo{author}{\bibfnamefont{S.~D.} \bibnamefont{Bartlett}},
  \bibinfo{author}{\bibfnamefont{L.~T.} \bibnamefont{Stephenson}},
  \bibnamefont{and} \bibinfo{author}{\bibfnamefont{B.~C.}
  \bibnamefont{Sanders}}, \emph{\bibinfo{title}{Quantum walks in higher
  dimensions}} (\bibinfo{year}{2001}), \eprint{quant-ph/0108004}.

\bibitem{tregenna02a}
\bibinfo{author}{\bibfnamefont{B.}~\bibnamefont{Tregenna}},
  \bibinfo{author}{\bibfnamefont{W.}~\bibnamefont{Flanagan}},
  \bibinfo{author}{\bibfnamefont{R.}~\bibnamefont{Maile}}, \bibnamefont{and}
  \bibinfo{author}{\bibfnamefont{V.}~\bibnamefont{Kendon}},
  \emph{\bibinfo{title}{Tuning quantum walks: coins and initial states}}
  (\bibinfo{year}{2002}), \bibinfo{note}{in preparation}.

\bibitem{kempe02a}
\bibinfo{author}{\bibfnamefont{J.}~\bibnamefont{Kempe}},
  \emph{\bibinfo{title}{Quantum random walks hit exponentially faster}}
  (\bibinfo{year}{2002}), \eprint{quant-ph/0205083}.

\bibitem{kendon02a}
\bibinfo{author}{\bibfnamefont{V.}~\bibnamefont{Kendon}},
  \bibnamefont{and} \bibinfo{author}{\bibfnamefont{B.}~\bibnamefont{Tregenna}},
  \emph{\bibinfo{title}{Decoherence is useful in quantum walks}}
  (\bibinfo{year}{2002}), \bibinfo{note}{submitted}, \eprint{quant-ph/0209005}.

\bibitem{brun02a}
\bibinfo{author}{\bibfnamefont{T.~A.} \bibnamefont{Brun}},
  \bibinfo{author}{\bibfnamefont{H.~A.} \bibnamefont{Carteret}},
  \bibnamefont{and} \bibinfo{author}{\bibfnamefont{A.}~\bibnamefont{Ambainis}},
  \emph{\bibinfo{title}{The quantum to classical transition for random walks}}
  (\bibinfo{year}{2002}), \eprint{quant-ph/0208195}.
\end{thebibliography}
\end{document}